\documentclass[twoside]{article}
\usepackage{fleqn}
\usepackage{espcrc2}

\usepackage[dvips]{color}

\usepackage{latexsym} % Gets \Box etc
\usepackage{amssymb}  % \gtrsim, \geqslant, etc etc: 
\usepackage{epsf}       % For PostScript figures
\usepackage{rotate}    % rotates PostScript figures

\newcommand{\al}{\alpha}
\newcommand{\be}{\beta}
\newcommand{\ga}{\gamma}

\newcommand{\si}{\sigma}

\newcommand{\<}{\langle} 
\renewcommand{\>}{\rangle} % LaTeX: \> already defined

\newcommand{\dsp}{\displaystyle}

\newcommand{\ad}{\dagger}
\newcommand\eqn[1]{(\ref{#1})}      % parentheses around the LaTex "ref" macro
\newcommand{\e}{ {\rm e} }
\newcommand{\beq}{\begin{equation}}
\newcommand{\eeq}{\end{equation}}
\newcommand{\ba}{\begin{array}}
\newcommand{\bea}{\begin{eqnarray}}
\newcommand{\ea}{\end{array}}
\newcommand{\eea}{\end{eqnarray}}

\newcommand{\nl}{\hfil\break}

\newcommand\comment[1]{ \hbox{[{\it Comment suppressed here.}\/]} }
\newcommand\hide[1]{}

\newcommand{\skipover}[1]{}

\def\phm{\phantom{-}}

\makeatletter %\catcode`\@=11

\newsavebox{\eqlabel}
\makeatletter  %\catcode`\@=11
\newlength{\numblen}
\newsavebox{\eqnumb}
\def\@eqnnum{\savebox{\eqnumb}{\rm (\theequation)}%
\settowidth{\numblen}{\usebox{\eqnumb}}%
\makebox[\numblen][l]{\usebox{\eqnumb}~~~\usebox{\eqlabel}}}
\makeatother   %\catcode`\@=12

\newenvironment{equationwithlabel}[1]{ %
  \begin{equation}\label{#1} }{\end{equation}} %\savebox{\eqlabel}{~}}
\newcommand{\beql}[1]{\begin{equationwithlabel}{#1}}
\newcommand{\eeql}{\end{equationwithlabel}}
\newcommand{\keV}{{\rm keV}}
\newcommand{\MeV}{{\rm MeV}}

\newcommand{\mf}{\rm}
\newcommand{\RR}{\mathbb{R}}

\newcommand{\upd}{{\rm update}}
\newcommand{\psibar}{\bar\psi}
\renewcommand{\sup}{\uparrow}
\newcommand{\sdn}{\downarrow}

\def\SU3{{\rm SU(3)}}
\def\U1{{\rm U(1)}}
\title{New possibilities for QCD at finite density}

\author{Mark Alford \address{School of Natural Sciences,
        Institute for Advanced Study, 
	Princeton, NJ 08540, USA\\
After Sept 1 1998: 
Center for Theoretical Physics, MIT, % Massachusetts Institute of Technology, 
Cambridge, MA 02139, USA}
}
       
\begin{document}

\begin{abstract}
I review the growing theoretical indications that at high densities
color SU(3) gauge symmetry is spontaneously broken by the formation of
a quark pair condensate.  This leads to a rich phase structure for QCD
as a function of temperature and chemical potential.

I also discuss the prospects for lattice QCD calculations
at finite density, including the Glasgow algorithm and
imaginary chemical potential.

\end{abstract}

\maketitle

\section{Superconducting phases of QCD}

The behavior of matter at high quark density is interesting in itself
and is relevant to phenomena in the early universe, in neutron stars,
and in heavy-ion collisions.
Recent analyses of Nambu--Jona-Lasinio (NJL) models of high-density QCD, 
in which the gluons are replaced by a four-fermion interaction,
indicate interesting physics. In particular,
exotic superconducting phases may occur above nuclear density.
Both two- and three-flavor cases have been studied, and I will
discuss them in turn, before going on to the topic of lattice techniques.

% I will review recent results suggesting
% a ``superconducting'' phase of QCD not far above nuclear density,
% describe the current status of the Glasgow method for lattice
% QCD at finite density, and discuss the possibility of using
% imaginary chemical potential in lattice simulations.

\subsection{Two-flavor QCD}

\begin{figure}[t]
\begin{center}
\epsfxsize=3.2in
\hspace*{0in}
%\vspace{-10cm}
%\epsffile{qcd2flav_print.eps}   % This looks better in black and white
 \epsffile{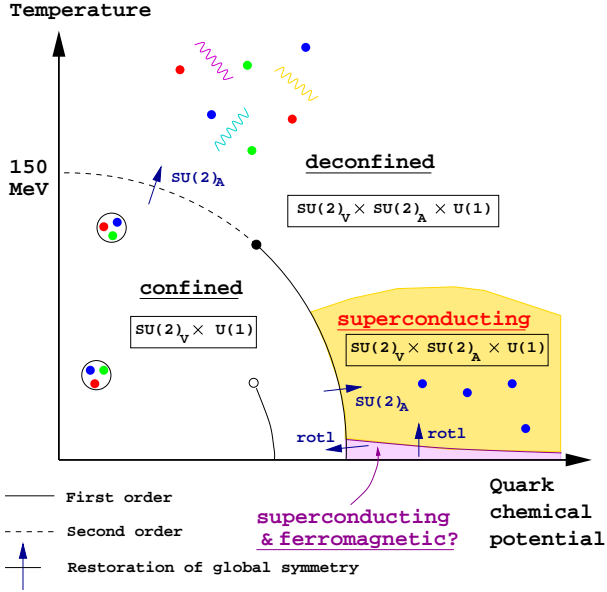}       % This looks better in color
\end{center}
\vspace{-6ex}
\caption{Conjectured phase diagram for QCD with two massless flavors.
Note the tricritical point (black dot), and exotic ``superconducting''
phases at high density and low temperature.
}
\label{fig:qcd2flav}
\end{figure}

Figure \ref{fig:qcd2flav} gives a plausible phase diagram for QCD with
two massless flavors. The main division is into chirally broken and
symmetric phases: axial flavor is broken at low
temperatures/densities, and restored at high ones. For a very clear
discussion and further references see Ref.~\cite{stonybr}. 
The chiral phase transition is believed to be second order at
$\mu=0$, and the indications from bag and matrix models are that it is
first order at $T=0$.
\looseness -1
% RM: Stephanov, PRL 76 (1996) 4472
Thus there is a tricritical point (solid circle) at the
switch-over from second order to first order \cite{br,stonybr}.
% When you turn on quark masses, the $2^{\mf nd}$ order chiral sym
% changing phase transitions become crossovers.
In the low-temperature chirally broken region there is the nuclear gas-liquid
phase transition line, which ends at a critical point (empty circle)
at $T\sim 10~\MeV$.

At low temperatures, it is becoming clear that additional interesting phases
occur above the chiral-symetry-restoring chemical potential.  It was originally
suggested by Bailin and Love \cite{bailin} (see also \cite{barrois})
that QCD at high density might behave analogously to a superconductor:
through the BCS mechanism \cite{bcs},
Cooper pairs of quarks condense in an attractive channel,
breaking the color gauge symmetry, and opening a gap at the Fermi
surface.  
% The two-flavor case:
Recent  mean-field/variational analyses of NJL models of QCD,
using the 4-leg instanton vertex as the effective interaction
\cite{arwtwo,instliq}, indicate that BCS-style 
quark pair condensation does indeed occur, and that
the gaps are phenomenologically significant---of order $100~\MeV$---%
at densities only a few times nuclear density.
The simplest form of pairing is a spin (indices suppressed) 
and flavor (indices $i,j$) singlet, which by antisymmetry of the
fermion wavefunctions must form a color $\bar{\bf 3}$ (indices $\al,\be$):
\beql{Cond2} 
\langle S| q_i^{\alpha} C \gamma^5 q_j^{\beta}|S\rangle 
~\propto ~\epsilon_{ij} \epsilon^{\alpha\beta 3}~,
\eeql

The coherent state $|S\>$, consisting of a quark pair condensate,
has lower free energy
than the perturbative vacuum, indicating that in the true vacuum
two quark colors (red and green, say)
condense, leaving the blue quarks forming
a Fermi surface. % (Fig.~\ref{fig:qcd2flav}).

In QCD, unlike the NJL model, color is a gauged symmetry, so
there is no local order parameter to distinguish the
superconducting phase of QCD from the deconfined one.
As in the standard model, however, we expect there to be a first-order
phase transition or crossover between regions of parameter space
with quite different physics.
The equation of state, according to NJL models, is only
slightly affected by quark pair condensation.  The two main features
that characterize the superconducting phase are

\noindent (1) There is a gap in the fermion spectrum.

\noindent (2) The condensate changes the electric charges of the quarks,
since it breaks color and electro\-magnetism
down to color $SU(2)$ and a new electromagnetism
that is a combination of the photon and one of the gluons. \nl
% \beq
% \ba{ll}
% \multicolumn{2}{l}{
% SU(3)_{\rm color}\times U(1)_Q \to SU(2)\times U(1)_{Q'},}\\[1ex]
% %{\rm where}\ Q' = Q + {\txt{1\over 6}} T_8; \\[1ex]
% Q'(u^{\rm red,green})= +\half, & Q'(u^{\rm blue})= +1,  \\[0.5ex]
% Q'(d^{\rm red,green})= -\half, & Q'(d^{\rm blue})= 0.
% \ea
% \eeq
It would be very useful to translate these into some observable for
heavy ion collisions, but this has not yet been done.

At very low temperatures, a more exotic phase may
form. The blue quarks, left out of the superconducting condensate,
may form spin-1 pairs and condense, breaking rotational invariance
and the remaining $SU(2)\times U(1)_{\rm em}$ gauge symmetry.
NJL calculations \cite{arwtwo} indicate a parameter-sensitive
and small gap ($\sim 1~\keV$) in this channel, leaving open
the intriguing possibility that such a phase might play a role in
neutron stars.

\begin{figure}[t]
\begin{center}
\epsfxsize=3.2in
\hspace*{0in}
\epsffile{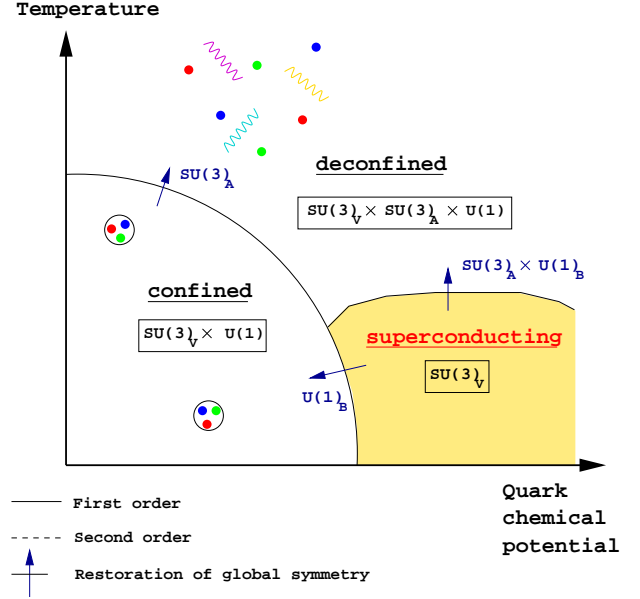}
\end{center}
\vspace{-6ex}
\caption{Conjectured phase diagram for QCD with three massless flavors.
Unbroken global symmetries are given in boxes.
The superconducting phase breaks baryon number and chirality.
}
\label{fig:qcd3flav}
\end{figure}

\subsection{Three-flavor QCD}

Fig.~\ref{fig:qcd3flav} shows a conjectured phase diagram for
QCD with 3 massless flavors \cite{arwthree}. 
We have to use the single-gluon-exchange vertex as the effective interaction,
since the instanton vertex now has an odd number of quark legs, 
and cannot be saturated by a quark pair condensate.
However, now that the number of flavors
and colors is the same, a different form of condensate is possible:
\beq
% {twoCondensates}
\<S| q^\alpha_i C\gamma^5  q^\beta_j |S\>
= \kappa_1 \delta^\alpha_i \delta^\beta_j 
+ \kappa_2 \delta^\alpha_j \delta^\beta_i~,
\eeq
The mixed Kronecker $\delta$ matrices are
invariant under correlated vectorial color/flavor
rotations (``color-flavor locking''). The breaking pattern is 
$\SU3_{\rm color}\times\SU3_L\times\SU3_R\times \U1_B \to \SU3_{\rm diag}$,
where $\SU3_{\rm diag}$ is the diagonal SU(3) subgroup 
of the first three factors.
The color symmetry (gauged in QCD) is broken by the quark pair condensate,
but, unlike the two-flavor case, chiral symmetry is also broken.
Also, with 3 flavors there {\em is} a gauge-invariant order parameter, 
corresponding to the breaking of baryon number: $\<NN\>$.

\subsection{Generalities}

\begin{figure}[htb]
\begin{center}
\epsfxsize=3.0in
\hspace*{0in}
\epsffile{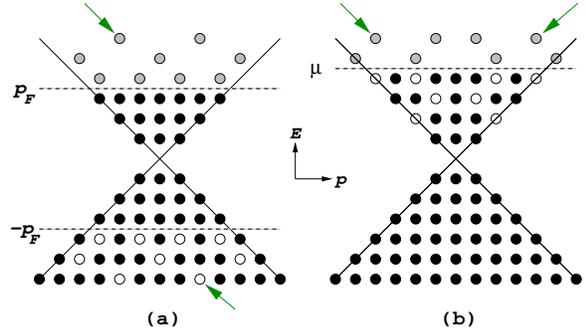}
\end{center}
\vspace{-6ex}
\caption{
Comparison of chiral and color symmetry breaking.
Shaded circles are particles, open circles are holes (antiparticles).
(a)~Chiral symmetry breaking: particles pair with antiparticles
of opposite momentum
(typical pair indicated by arrows).
(b)~Color symmetry breaking: particles pair with particles of
opposite momentum (typical pair indicated by arrows), 
antiparticles with antiparticles.
}
\label{fig:fermisea}
\end{figure}

The detailed NJL-model calculations that show quark pair condensation
are given in \cite{arwtwo,instliq,arwthree}. 
% Here I just want to give a
% qualitative comparison of chiral and color symmetry breaking.
Nambu and Jona-Lasinio explicitly based their studies of chiral
symmetry breaking on BCS, and described the chiral condensate as a
"'superconductive' solution" \cite{njl}. However, there are important
differences (Fig.~\ref{fig:fermisea}).  Chiral symmetry breaking is
caused by a condensate of particle-antiparticle pairs with
zero net momentum. In the presence of a Fermi surface with Fermi
momentum $p_F$, one can only create particles with $p>p_F$, so as the
density grows, more and more states are excluded from pairing, and
chiral symmetry breaking is suppressed.  In contrast, color symmetry
breaking involves pairs of particles or pairs of antiparticles. Near
the Fermi surface these pairs can be created at negligible cost in
free energy, and so any attractive particle-particle
interaction enables the pairs to lower the free
energy. This is the BCS instability of the perturbative vacuum.  If
there is any channel in which the interaction between quarks is
attractive, then quark pair condensation in that channel will
occur. As density increases, the phase space available near the Fermi
surface grows, and more quark pairing occurs.

There are many directions in which to continue investigating
quark pair condensates:
% \looseness=-1
experimental signatures in heavy ion collisions,
their possible role in neutron star physics,
and of course
the 2+1 flavor case, with a realistic strange quark mass.

Finally, it would be valuable to perform lattice calculations
of the finite density behavior of both the toy models and QCD.
Hands and Morrison \cite{HM.Jul98} have performed lattice simulations
of a Gross-Neveu model, which is very similar to the NJL model
used here, but they have not yet seen unambiguous evidence
of quark pairing. Finite density calculations in QCD, however,
remain a much more elusive goal. This is the topic of the rest of
this paper.

%%%%%%%%%%%%%%%%%%%%%%%%%%%%%%%%%%%%%%%%%%%%%%%%%%%%%%%%%%%%%%%%%%%%%%%
%                                                                     %
%                                                                     %
%%%%%%%%%%%%%%%%%%%%%%%%%%%%%%%%%%%%%%%%%%%%%%%%%%%%%%%%%%%%%%%%%%%%%%%

\section{Lattice QCD at finite density}

The usual approach to fermions is to integrate them out:
% \beql{latt:Z}
% \ba{rcl}
% Z(\mu) &=&  \quad\, 
% \dsp \sum_N Z_N \e^{-\mu N} \\[3ex]
%  &=& {\dsp \sum_{U(x) {\mf ~configs}} }
%    \underbrace{\det M \,\e^{-S_{\mf gauge}[U]} }_{\mf sampling~weight} \\[4ex]
% S_{\mf ferm} &=& \int_x \psibar M \psi 
% \ea
% \eeql
\beql{latt:Z}
\ba{r@{~}c@{~}l}
Z(\mu) &=&  \dsp \sum_N Z_N \e^{-\mu N} = \!\!\!\!\!\!
{\dsp \sum_{\scriptsize \ba{c} U(x) \\ {\mf ~configs}\ea} }
 \!\!\! 
 \underbrace{\det M \,\e^{-S_{\mf glue}[U]} }_{\mf sampling~weight} \\[4ex]
S_{\mf ferm} &=& \int_x \psibar M \psi 
\ea
\eeql

For Monte Carlo evaluation,
% of the partition function \eqn{latt:Z},
the sampling weight must be positive, so $\det M$ must be
non-negative for any gauge configuration. 
\looseness=-1
One way to guarantee this is if we have an even number of flavors,
each with the same fermion matrix $M$,
and $M$ is similar to its adjoint, so 
the eigenvalues are real or in complex-conjugate pairs,
and $\det M \in\RR$.
\beql{latt:Mreal}
M^\ad = P M P^{-1} \hbox{~for some~}P,\quad 
N_F \hbox{~even}
\eeql
% So at $\mu=0$, $M^\ad = \ga_5 M \ga_5$, \\
% so $\det M$ is real (positive for even $N_F$).
This is the situation in zero-density lattice
QCD. For the Wilson action, for example,
\beq
\ba{rcl}
M &=& \phm\ga^\mu D_\mu + rD^2 + m + \mu\ga_0  \\[1ex]
 M^\ad &=& -\ga^\mu D_\mu + rD^2 + m + \mu\ga_0
\ea
\eeq
so without a chemical potential \eqn{latt:Mreal}
is obeyed with $P=\ga_5$, but introducing a real
chemical potential violates this condition.
$\det M$ is then complex, and 
\looseness=-1
straightforward Monte-Carlo methods are inapplicable.
Similar conclusions obtain for Kogut-Susskind quarks.
This is the ``sign problem'', which is really a phase problem for QCD.
It is interesting to note, however, that an {\em imaginary}
chemical potential leaves the measure positive.

%Mod det is used by the Japanese (Iwasaki? Kanawa?) for
%odd $N_F$ at finite temp.

\subsection{The Glasgow method}
For the last decade, the approach to finite density QCD that has been
most seriously pursued is the Glasgow method,
\cite{originalBarbour,Glasgow97}. 
This avoids the positivity problem by treating the
chemical potential analytically: $Z$ is expanded
in powers of the fugacity $\e^{\be\mu}$, and the coefficients
are evaluated by Monte-Carlo using the $\mu=0$ weighted ensemble.
\beql{Glasgow:ratio}
 Z(\mu) = \left\< \det M(\mu) \over \det M(0) \right\>_{\mu=0}
\eeql
Unfortunately, the Glasgow method does not reproduce the most
fundamental property expected of QCD at finite chemical potential,
namely the onset of baryon density when $\mu$ reaches $M_{\rm baryon}/3$.
Instead, the onset seems to begin at a lower chemical potential
$\mu_o \sim M_\pi/2$.
This is the behavior of the quenched theory, in which there is
an unphysical baryonic pion state \cite{BaryonicPion}.

%Signs of critical behavior at the expected $\mu_c = m_N/3$. \\
%Disputed as numerical artefact, \\[-1ex]
%{\scriptsize Aloisio, Azcoiti, Di Carlo, Galante, Grillo,
% \hspace{1em}{\tt hep-lat/9807003}}

%\item Heavy quarks. \hspace{1em}
%{\scriptsize Blum, Hetrick, Toussaint %{\tt hep-lat/9802004}
%}

% Azcoiti et al argue that one must ensure that the expansion coeffs
% of GCPF are symmetric around central coeff.

%\vspace{0.5cm}
\begin{figure}[tb]
\begin{center}
\epsfxsize=2.5in
\hspace*{0in}
\rotate[r]{ \epsffile{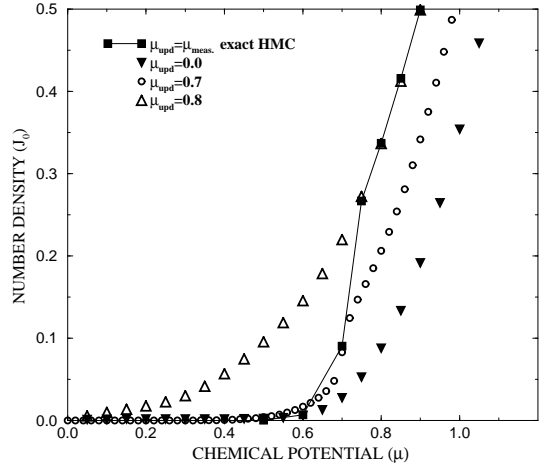} }
\end{center}
\vspace{-3ex}
\caption{
Fermion density as a function of chemical potential $\mu$ for
the Gross-Neveu model \protect\cite{MH.Biele98},
calculated from moderate statistics using Glasgow reweighting
of ensembles generated at chemical potentials $\mu_\upd=
0.0,0.7,0.8$, as well as the exact (unreweighted) result.
Note how the apparent
onset depends on $\mu_\upd$, indicating that the
$\mu_\upd=0$ ensemble used for QCD may give unreliable
results at moderate statistics.
}
\label{fig:GNonset}
\end{figure}

In the last year, there has been a convergence of opinion that the
Glasgow method will eventually show the correct onset behavior, but
only at very large statistics.  This is because of the ``overlap
problem'', a mismatch between the regions of configuration space that
are emphasized by the $\mu=0$ ensemble, and the regions that are
relevant for the finite-density physics. Because of the
measure mismatch, the integral comes from
large but rare fluctuations in the reweighting factor in
\eqn{Glasgow:ratio}, making it necessary to accumulate a huge number
of configurations.

The evidence for this conclusion has come from
several sources.
(1)~Using the Glasgow method for
QCD on a tiny $2^4$ lattice, Barbour \cite{IB.Biele98}
finds that $\mu_{\mf o}$ rises towards $m_N/3$ at 
very high statistics.
(2)~The Gross-Neveu model has been studied
in 2+1 dimensions \cite{MH.Biele98}.
It has quarks and conjugate quarks, so there is
no sign problem, and one can perform updates
at any chemical potential. At moderate statistics,
the onset behavior shows dependence
on the value of $\mu$ used in the updating (see Fig.~\ref{fig:GNonset}).
(3)~Azcoiti et al \cite{VAetal} argue that
the phase of the quark determinant
drops as $\exp(-V)$, so statistics $ \sim \exp(V)$ is needed.
(4)~Halasz \cite{MH} has studied matrix models of
the QCD fermion determinant,
and finds that they need statistics $\sim \exp(N)$.

%%%%%%%%%%%%%%%%%%%%%%%%%%%%%%%%%%%%%%%%%%%%%%%%%%%%%%%%%%%%%%%%%%%%%%%
%                                                                     %
%                                                                     %
%%%%%%%%%%%%%%%%%%%%%%%%%%%%%%%%%%%%%%%%%%%%%%%%%%%%%%%%%%%%%%%%%%%%%%%

\subsection{Imaginary chemical potential}
As was noted above,
with imaginary chemical potential $\mu = i\nu$,
$\det M$ is real, and the functional integral
can be evaluated by standard Monte-Carlo methods \cite{Sugar,ht,AKW}.
We still have to choose an updating value $\nu_\upd$, and then
``reweight'' to obtain $Z(i\nu)$, but the reweighting factor
is always positive, so there is no phase problem,
and very high statistics is not needed.
\beq
{Z(i\nu)\over Z(i\nu_\upd)}
 = \left\< \det M(i\nu) \over \det M(i\nu_\upd) \right\>_{\mu=i\nu_\upd}
\eeq
We can cover $\nu=0\ldots 2\pi/\beta$ with ``patches'' centered
at several different $\nu_\upd$, and thereby ensure that the 
reweighting factor has arbitrarily small fluctuations.
Then we must Fourier transform to get canonical partition functions
\beql{imag:ft}
Z_N=\frac{\beta}{2\pi} \int_0^{2\pi/\beta}\!\! d\nu 
  \, Z(i\nu) \e^{-i\beta\nu N}.
\eeql

The Fourier transform is the place where large errors may arise.
Imaginary chemical potential does not bias the ensemble towards states
with higher quark number; it relies on thermal or quantum fluctuations
to supply such states, and it gives them a characteristic weighting.
If such fluctuations are too rare then $Z(i\nu)$ will
\looseness=-1
be dominated by $Z_0$, and very large statistics
will be needed to see the effects of the higher $Z_N$. 
This is not necessarily a problem: as the temperature $T$ rises
towards the deconfining phase transition $T_c$, the baryon
becomes lighter, so $M_B/T$ may become small enough for
thermal fluctuations to populate the system with baryons.
Calculating the $Z_N$ by \eqn{imag:ft} will then be straightforward.
The Glasgow algorithm may also be expected to work better as $T\to T_c$.

\subsection{The Hubbard model with imaginary $\mu$}

\begin{figure}[tb]
\begin{center}
\epsfxsize=2.5in
\hspace*{0in}
\epsffile{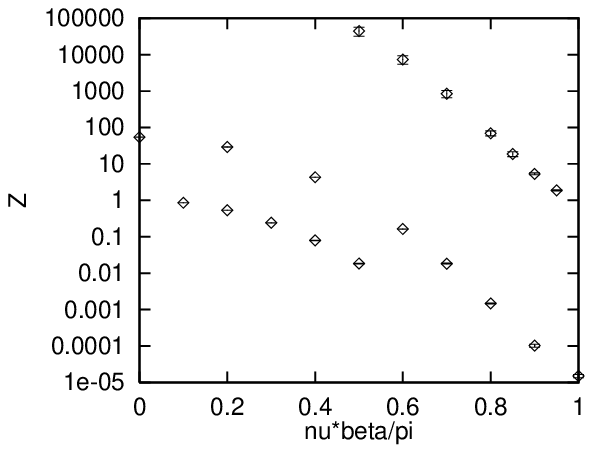}
\end{center}
\begin{center}
\epsfxsize=2.5in
\hspace*{0in}
\epsffile{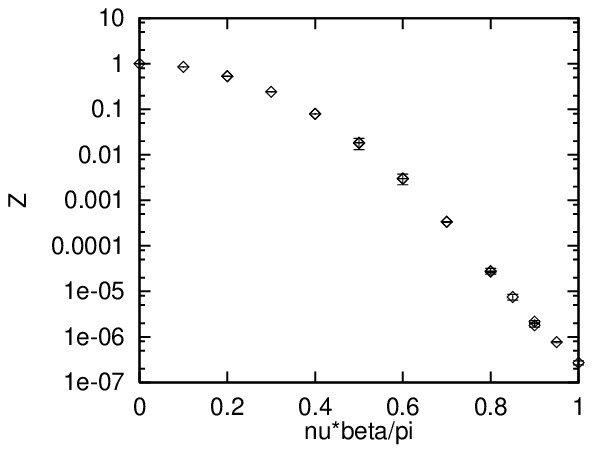}
\end{center}
\vspace{-3ex}
\caption{
$Z(i\nu)$ for the Hubbard model
on a $4^2\times 10$ lattice, with $K=1,\ \be=1.5,\ U=1.0$ \cite{AKW}.
The three patches agree well when given the same normalization.
}
\label{fig:HubZnu}
\end{figure}

As a test of the practicality of imaginary chemical potential,
it is interesting to explore a simple theory that
has the sign problem at non-zero chemical potential.
The natural candidate is the Hubbard model in 2 dimensions
\cite{Sugar,AKW},
% non-relativistic electrons on a lattice
% with a hopping term and on-site repulsion.
\beq
\ba{r@{~}c@{~}l}
{\cal H} &= &
- K \sum_{\<i,j\>,\si} a^\ad_{i\si} a^{\phantom\ad}_{j\si} \\[2ex]
&& - {U\over 2} \sum_i (a^\ad_{i\sup} a^{\phantom\ad}_{i\sdn}
     -a^\ad_{i\sdn} a^{\phantom\ad}_{i\sup} )^2
+ \mu  \sum_{i,\si}  a^\ad_{i\si} a^{\phantom\ad}_{i\si}
\ea
\eeq
By a particle-hole transformation, $\mu=0$ gives half-filling.
We can replace the four-fermion interaction with an
auxiliary field $A$,
\beq
Z(\mu) = \sum_{A(x)} \e^{-A^2/2} \det M(\mu) \, \det M(-\mu).
\eeq
The matrices $M$, given in \cite{Creutz}, are real for
real $\mu$, so the sampling weight
is real but not positive, and Monte-Carlo is impossible.
For $\mu$ zero or imaginary, the weight is $|\det M|^2$, and is positive, so
we can calculate ratios of partition functions
\[
{ Z(i\nu)\over Z(i\nu_0)} = \sum_{A(x)} \e^{-A^2/2} \det |M(i\nu_0)|^2 \, 
{ |\det M(i\nu)|^2 \over |\det M(i\nu_0)|^2 },
\]
and use several values of $\nu_\upd=\nu_0$ to 
eliminate measure-mismatch. In Fig.~\ref{fig:HubZnu} we show results for
$Z(i\nu)/ Z(i\nu_0)$ using three values of $\nu_0$. Combining them
to obtain a single plot of $Z(i\nu)/ Z(0)$, we see that that the
three patches agree to within their statistical errors.

Finally, we fit our $Z(i\nu)$ data to $\exp(-a\nu^2)\times{\mf spline}$, 
and Fourier transform it to obtain $Z_N$ (Fig.~\ref{fig:HubZnu}).  At
this relatively high temperature we have no trouble obtaining $Z_N$ up
to $N=5$. It would be interesting to see whether similar results could
be obtained for QCD at temperatures just below the phase transition.

% 2D Hubbard model is a candidate for describing high-$T_c$
% superconductors. Maybe at some filling fraction the
% electrons form Cooper pairs and condense, to make a
% superconductor?
% 
% No sign of pairing, but we got $Z_N$ up to $N=5$, using embarrassingly
% inefficient updating algorithm: Mathematics ``Det'' function.

\begin{figure}[tb]
\begin{center}
\epsfxsize=3in
\epsffile{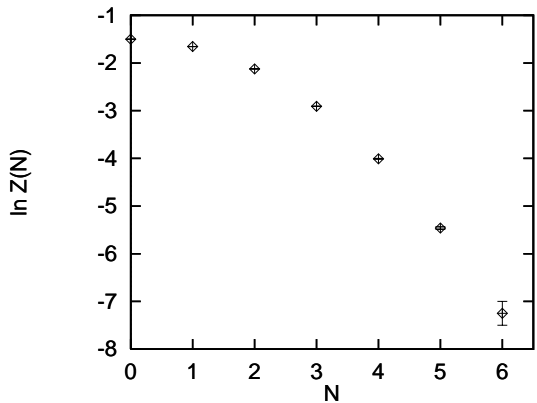}
\end{center}
\vspace{-10ex}
\caption{
$Z_N$ for the Hubbard model, obtained by inverse Fourier transform
of the data in Fig.~\ref{fig:HubZnu}.
}
\vspace{-3ex}
\label{fig:HubZN}
\end{figure}

\section{Conclusions}

In summary, while there are growing indications 
from Nambu--Jona-Lasinio and other models of exotic
phenomena in high-density QCD, it remains very difficult
to perform the necessary lattice calculations.
% quark pair condensation, color superconductivity, modified electromagnetism,
% baryon number violation. \\
% Heavy ion collisions, supernovas, neutron stars$\ldots$

The Glasgow algorithm does not appear to be practical, since even on a
$4^4$ lattice it needs more statistics than anyone has been able to
gather.  Imaginary chemical potential is an interesting alternative,
since it is quite possible that the numerical problems of inverse
Fourier-transforming $Z(i\nu)$ to $Z_N$ will be less difficult.
% than overcoming the Glasgow algorithm's mismatch of measure.

Several other approaches are under development, including perfect
actions \cite{Bietenholtz}, computer evaluation of the fermionic
Grassman integrals \cite{grassman}, and projecting out the
$N$-particle sectors using a new dynamical fermion algorithm
\cite{lls}.  Though it is proving hard to find, a successful approach
will open up new areas of phenomenology to lattice gauge
theorists. Clearly the rewards will be worth the effort.

% Mendel, hep-lat/9807004, suggests suppressing flavor-changing
% hard gluons will help.
%

\vspace{1ex}
% {\samepage 
% \begin{center} Acknowledgements \end{center}
% \nopagebreak
% This work was supported by DOE grant DE-FG02-90ER40542,
% and by the generosity of Frank and Peggy Taplin.
% \par
% }
\noindent
{\bf Acknowledgements:}
This work was supported by DOE grants DE-FG02-90ER40542
and DF-FC02-94ER40818,
and by the generosity of Frank and Peggy Taplin.

\end{document}